\documentclass[10pt,twocolumn,english,secnumarabic,nofootinbib,aps,superscriptaddress,preprintnumbers,prd,bibnotes]{revtex4-2}
\pdfoutput=1

\PassOptionsToPackage{linecolor=blue,backgroundcolor=blue!25,bordercolor=blue,textsize=scriptsize}{todonotes}
\usepackage[utf8]{inputenc}
\usepackage{babel}
\usepackage{verbatim}
\usepackage{refstyle}
\usepackage{mathtools}
\usepackage{todonotes}
\usepackage{amsmath}
\usepackage{amsthm}
\usepackage{amssymb}
\usepackage{graphicx}
\usepackage{wasysym}
\usepackage{microtype}
\usepackage[unicode=true,
bookmarks=true,bookmarksnumbered=false,bookmarksopen=false,
breaklinks=false,pdfborder={0 0 0},pdfborderstyle={},backref=false,colorlinks=true]
{hyperref}
\hypersetup{pdftitle={Calabi--Yau Metrics, Energy Functionals and Machine-Learning},
	pdfauthor={},
	linkcolor=darkblue, urlcolor=darkblue, citecolor=darkblue, filecolor=black, linktoc=page}

\makeatletter

\AtBeginDocument{\providecommand\eqref[1]{\ref{eq:#1}}}
\RS@ifundefined{subsecref}
  {\newref{subsec}{name = \RSsectxt}}
  {}
\RS@ifundefined{thmref}
  {\def\RSthmtxt{theorem~}\newref{thm}{name = \RSthmtxt}}
  {}
\RS@ifundefined{lemref}
  {\def\RSlemtxt{lemma~}\newref{lem}{name = \RSlemtxt}}
  {}

\pdfoutput=1

\usepackage{color}
\usepackage{babel}
\usepackage{amsthm}
\usepackage{setspace}

\usepackage[]{todonotes}

\AtBeginDocument{}
\RS@ifundefined{subsecref}{\newref{subsec}{name = \RSsectxt}}{}
\RS@ifundefined{thmref}{\def\RSthmtxt{theorem~}\newref{thm}{name = \RSthmtxt}}{}
\RS@ifundefined{lemref}{\def\RSlemtxt{lemma~}\newref{lem}{name = \RSlemtxt}}{}

\usepackage{tikz}
\usetikzlibrary{backgrounds}
\usepackage{slashed}
\usepackage{amsfonts}
\SetTracking{encoding={*}, shape=sc}{0} 	
\usepackage{color}
\definecolor{darkblue}{rgb}{0.0,0.0,0.4} 	
\definecolor{darkred}{rgb}{0.4,0.0,0.0} 	

\allowdisplaybreaks		

\g@addto@macro\bfseries{\boldmath}

\g@addto@macro\@floatboxreset\centering

\let\originalleft\left
\let\originalright\right
\renewcommand{\left}{\mathopen{}\mathclose\bgroup\originalleft}
\renewcommand{\right}{\aftergroup\egroup\originalright}

\usepackage{tikz}
\usetikzlibrary{shapes.geometric,arrows} 

\tikzstyle{block} = [rectangle, draw,    
text width=15em, text centered, inner ysep = 1em, minimum height=1em] 
\tikzstyle{netblock} = [rectangle, draw, color=darkblue,   
text width=10em, text centered, inner ysep = 1em, minimum height=1em]
\tikzstyle{netblock2} = [rectangle, draw, color=darkred,  rounded corners,   
text width=10em, text centered, inner ysep = 1em, minimum height=1em ] 
\tikzstyle{netblock3} = [rectangle, draw, color=darkblue,   
text width=8em, text centered, inner ysep = 1em, minimum height=1em ]
\tikzstyle{netblock4} = [rectangle, draw, color=darkblue,   
text width=3.5em, text centered, inner ysep = 0.0em, minimum height=1.5em ]
\tikzstyle{line} = [draw, -latex']

\tikzset{every picture/.style={line width=0.75pt}} 

\makeatother

\begin{document}
\preprint{LIMS-2021-018}
\title{Calabi--Yau Metrics, Energy Functionals and Machine-Learning}
\author{Anthony Ashmore}
\email[]{ashmore@uchicago.edu}

\affiliation{Enrico Fermi Institute \& Kadanoff Center for Theoretical Physics, University of Chicago, IL 60637, USA}
\affiliation{Sorbonne Université, Laboratoire de Physique Théorique et Hautes Energies, F-75005 Paris, France}
\author{Lucille Calmon}
\email[]{m.l.calmon@qmul.ac.uk}

\affiliation{School of Mathematical Sciences, Queen Mary, University of London, London E1 4NS, UK}
\affiliation{Department of Physics, Imperial College London, Prince Consort Road, London, SW7 2AZ, UK}
\author{Yang-Hui He}
\email[]{hey@maths.ox.ac.uk}

\affiliation{London Institute for Mathematical Sciences, Royal Institution, W1S 4BS, UK}
\affiliation{Department of Mathematics, City, University of London, EC1V0HB, UK}
\affiliation{Merton College, University of Oxford, OX1 4JD, UK}
\affiliation{School of Physics, NanKai University, Tianjin, 300071, P.R.~China}
\author{Burt A.~Ovrut}
\email[]{ovrut@elcapitan.hep.upenn.edu}

\affiliation{Department of Physics, University of Pennsylvania, Philadelphia, PA 19104, USA}
\begin{abstract}
\noindent We apply machine learning to the problem of finding numerical Calabi--Yau metrics. We extend previous work on learning approximate Ricci-flat metrics calculated using Donaldson's algorithm to the much more accurate ``optimal'' metrics of Headrick and Nassar. We show that machine learning is able to predict the K\"ahler potential of a Calabi--Yau metric having seen only a small sample of training data.
\end{abstract}
\maketitle
\global\long\def\KP{\mathcal{K}}%
\global\long\def\bP{\mathbb{P}}%
\global\long\def\op#1{\operatorname{#1}}%
\global\long\def\vol{\op{vol}}%
\global\long\def\Vol{\op{Vol}}%
\global\long\def\dd{\text{d}}%
\global\long\def\ii{\text{i}}%
\global\long\def\ee{\text{e}}%
\global\long\def\hpsi{\hat{\psi}}%
\global\long\def\Ex#1{\text{E}_{#1}}%

\section{Introduction \& summary}

\noindent Recent decades have seen great progress in trying to match string theory to experiment. There are now large classes of top-down string models with the potential to reproduce the Standard Model at low energies. Perhaps the most promising of these approaches is also the oldest, namely compactifying the $\Ex 8\times\Ex 8$ heterotic string on a six-dimensional Calabi--Yau manifold (or on a threefold times an interval in heterotic M-theory \cite{Horava:1995qa,Horava:1996ma,Lukas:1997fg, Lukas:1998yy, Lukas:1998tt}). Within this framework, there are vast numbers of models which give minimally supersymmetric extensions of the Standard Model, including three generations, the proper Higgs structure, and the correct gauge group~\cite{Braun:2005nv,Braun:2005bw,Braun:2005ux,Bouchard:2005ag,Anderson:2009mh,Braun:2011ni,Anderson:2011ns,Anderson:2012yf,Anderson:2013xka,Nibbelink:2015ixa,Nibbelink:2015vha,Braun:2017feb,Constantin:2018xkj}.

Unfortunately, in nearly all of these models, we are currently unable to make sharp predictions for the couplings and masses that govern the resulting four-dimensional effective physics. This can be traced to our ignorance of the explicit Calabi--Yau metric on the compactification manifold (and an explicit hermitian Yang--Mills connection for the gauge fields). Without this metric, we are generally unable to compute the superpotential or the Kähler potential of the effective theories, nor fix a supersymmetry-breaking mechanism, and hence unable to compare predictions with experiment.

The recent years have seen a flurry of activity in tackling the problem of finding Calabi--Yau metrics by turning to numerical methods and, most recently, machine learning~\cite{He:2017aed,He:2017set,Krefl:2017yox,Carifio:2017bov,Ruehle:2017mzq} (see \cite{He:2018jtw,He:2019nzx,Alessandretti:2019jbs,He:2021oav} for some uses of ML in pure mathematics). 
There are now a variety of methods for computing these metrics numerically, including position space methods~\cite{Headrick:2005ch}, spectral methods~\cite{Tian,math/0512625,Douglas:2006rr,Braun:2007sn,Headrick:2009jz}, and machine learning algorithms~\cite{Ashmore:2019wzb,Anderson:2020hux,Douglas:2020hpv,Jejjala:2020wcc,Douglas:2021zdn,Larfors:2021pbb}. 

This note completes on the work started in \cite{Ashmore:2019wzb}. There, three of the present authors attempted to understand the extent to which machine-learning techniques could be used to speed up or improve the accuracy of numerically calculated Calabi--Yau metrics. In that work, the focus was on Ricci-flat metrics computed using Donaldson's balanced metrics~\cite{math/0512625}. A combination of point-wise extrapolation and gradient-boosted decision trees was used to improve the accuracy of these numerical metrics and to reduce the total time taken to calculate them. Here, we extend and complete this work by showing that a neural network can also learn the much more accurate ``optimal'' metrics, first discussed by Headrick and Nassar~\cite{Headrick:2009jz}. These metrics are exponentially more accurate than those found via balanced metrics.

For concreteness, we focus on perhaps the most famous Calabi--Yau manifold: the Fermat quintic. This threefold admits a large discrete symmetry group, $S_{5}\ltimes(\mathbb{Z}_{5})^{4}$, of order 150,000, which can be used to reduce the basis of polynomials and hence the space of Kähler potentials that are minimised over. We use a five-layer neural network to encode the Kähler potential of the approximate Ricci-flat metric. The network is trained via supervised learning, with inputs given by points on the Calabi--Yau hypersurface and the values of certain sections, and the outputs given by the numerical value of the exponential of the Kähler potential at those points. The outputs are calculated using the \texttt{Mathematica} package \texttt{fermat.m} available at \cite{fermat.m}. 

We find that this neural network can mimic the behaviour of the Kähler potential to high accuracy, having been trained on approximately 2000 input-output pairs. We show that the performance of the network is robust by calculating a ten-run averaged loss measure and showing that the accuracy does not vary significantly between each run. We also examine how the accuracy of the network varies as a function of training set size, showing that the mean absolute percentage error between the network outputs and the target Kähler potential decreases as the training fraction increases.

\section{Background and notation}

\noindent For the purposes of this work, a Calabi--Yau (CY) manifold $X$ is a compact, Kähler manifold which admits a Ricci-flat metric. Thanks to Yau's proof~\cite{Yau:420951} of the Calabi conjecture~\cite{Calabi57}, we know that such metrics exist provided the first Chern class $c_{1}(X)$ of $X$ vanishes. We let $x_{a}$, $a=1,\ldots,3$, denote complex coordinates on a threefold $X$. We will focus on the example of the Fermat quintic defined by the zero locus in $\mathbb{P}^{4}$ of the equation 
\begin{equation}
Q\equiv z_{0}^{5}+z_{1}^{5}+z_{2}^{5}+z_{3}^{5}+z_{4}^{5}=0,\label{eq:fermat}
\end{equation}
where the $z_{i}$, $i=0,\dots,4$ are homogeneous coordinates on projective space. Since a Calabi--Yau manifold is Kähler and the holomorphic $(3,0)$-form is determined exactly by $Q$ via a residue theorem, the problem of finding the Ricci-flat metric reduces to finding a suitable Kähler potential, $\KP$.

\subsection{The energy functional method}

\noindent The energy functional method was developed by Headrick and Nassar in \cite{Headrick:2009jz} with the aim of computing numerical Ricci-flat metrics on Calabi--Yau manifolds given by hypersurfaces in projective space (or products thereof).\footnote{Further details can be found in their unpublished notes available at \cite{fermat.m}.} We now review some of the important steps.

First, one parametrises the space of Kähler potentials using the ``algebraic metrics'' ansatz of Tian~\cite{Tian} and Donaldson~\cite{math/0512625}. In our case, $\KP$ is written as an expansion in the eigenfunctions of the Laplacian on $\bP^{4}$. This provides a controlled expansion in terms of a parameter $k$ when one includes eigenfunctions for the first $k+1$ eigenspaces. Explicitly, the first $k$ eigenspaces of the Laplacian are spanned by
\begin{equation}
\frac{s_{\alpha}\bar{s}_{\bar{\beta}}}{(\delta^{i\bar{j}}z_{i}\bar{z}_{\bar{j}})^{k}},
\label{eq:eigenspaces}
\end{equation}
where the $s_{\alpha}(z)$ are homogeneous polynomials of degree $k$ in the $z_{i}$, and $\delta^{i\bar{j}}$ is the hermitian form which gives the Fubini--Study metric on $\bP^{4}$.

Since $Q=0$ on the hypersurface, one should quotient the set of homogeneous polynomials by the ideal generated by $Q$. This gives a reduced basis, which we denote by $p_{A}$. Any Kähler metric with potential $\KP$ in the same Kähler class as the Fubini--Study metric differs from $\mathcal{K}_{\text{FS}}$ by a globally defined function. Introducing coordinates $u$ on the patch $O_{(c)}\subset X$ (defined by $z_{c}\neq0$), one can expand the Kähler potential in the reduced $p_{A}$ basis as
\begin{equation}
\KP=\frac{1}{k}\ln\left(h^{A\bar{B}}p_{A}\bar{p}_{\bar{B}}\right)=\frac{1}{k}\ln\psi,\label{eq: algebraic kahler pot}
\end{equation}
where $\psi\equiv h^{A\bar{B}}p_{A}\bar{p}_{\bar{B}}$ and the explicit form of $\mathcal{K}_{\text{FS}}$ was used. The resulting metrics, which depend on the parameters $h^{A\bar{B}}$, are known as ``algebraic metrics''.

If the zero locus of $Q$ is invariant under the action of a group, the Ricci-flat Kähler potential and the metric itself must also be invariant, and so the group gives an isometry. In practice, this means we can restrict the combinations of $p_{A}\bar{p}_{\bar{B}}$ that appear in $\KP$ to be those invariant under the isometry group. The Fermat quintic \eqref{fermat} admits a discrete $S_{5}\ltimes(\mathbb{Z}_{5})^{4}$ symmetry, which can be thought of as permutations of the $z_{i}$ combined with phase rotations.\footnote{A discussion of the generators of this group can be found in \cite{Douglas:2006rr,Braun:2008jp,Ashmore:2021qdf}.} This prompts the definition of a new basis spanned by polynomials of $u$ and $\bar{u}$:
\begin{equation}
\mathcal{P}_{l}=c_{l}^{I\bar{J}}\rho_{I}\bar{\rho}_{\bar{J}},\label{eq: vector basis}
\end{equation}
where the $\rho_I$ are, like the $p_A$, functions of $u$. Here, the coefficients $c_{l}^{I\bar{J}}$ pick out the polynomials that are linearly independent on the hypersurface $Q=0$ and invariant under the discrete isometry group. In this basis, the function $\psi$ is simply $\psi=h^{l}\mathcal{P}_{l}$, with the resulting Kähler potential $\KP[h]$ given as a function of $h^{l}$. The coefficients $h^{l}$ are the adjustable parameters that one can tweak to find the best approximation to the Ricci-flat metric for a choice of degree $k$.

In \cite{Headrick:2009jz}, the problem of finding an approximate Ricci-flat metric on a Calabi--Yau manifold was rephrased as the minimisation of an appropriate functional. Recall that there are two natural volume forms on a Calabi--Yau threefold, defined by the holomorphic $(3,0)$-form and the Kähler metric:
\begin{equation}
\vol_{\Omega}=\ii\,\Omega\wedge\bar{\Omega},\qquad\vol_{\omega}=\omega\wedge \omega\wedge \omega.\label{eq:volumes}
\end{equation}
The first, $\vol_{\Omega}$, depends only on the defining equation $Q=0$, while the second depends on the explicit choice of Kähler potential. Defining the ratio of these as
\begin{equation}
v_{\omega}\equiv\frac{\vol_{\omega}}{\vol_{\Omega}},\label{eq: ration omega start}
\end{equation}
the Ricci tensor of the Kähler metric is given by
\begin{equation}
\mathcal{R}_{a\bar{b}}=-\partial_{a}\bar{\partial}_{\bar{b}}\ln v_{\omega}.
\end{equation}
Yau's theorem then ensures that there is a unique choice of $\omega$ for which the Ricci tensor vanishes, with the Ricci-flat metric lying in the same Kähler class as the original Fubini--Study metric. From above, the vanishing of the Ricci tensor is equivalent to $v_{\omega}=\text{constant}$. Without loss of generality, the coefficients in (\ref{eq:volumes}) can be chosen so that $v_{\omega}=1$ for the Ricci-flat representative.

The functional chosen in \cite{Headrick:2009jz} was
\begin{equation}
E[\omega]=\int_{X}\vol_{\Omega}(1-v_{\omega})^{2}.\label{functional}
\end{equation}
Importantly, $E[\omega]$ is non-negative, has a unique minimum on the Ricci-flat metric, and has no other critical points. Thinking of the functional as $E[h]$, i.e.~a functional of the parameters of the underlying Kähler potential, an approximate Ricci-flat Kähler potential can be found by minimising $E[h]$ over the parameters $h^{l}$.\footnote{Note that for arbitrary Kähler potentials, $E[\KP]$ does not have a unique minimum since Kähler transformations change $\KP$ but leave $\omega$ (and hence $E[\KP]$) unchanged. For Kähler potentials defined via $\psi=h^{l}\mathcal{P}_{l}$, changing the $h^{l}$ never corresponds to a Kähler transformation, and so $E[h]$ does have a unique minimum.}

The ratio $v_{\omega}$ can be rewritten to make its dependence on the parameters $h^{l}$ explicit and allow efficient numerical computation. Upon defining the four-component object
\begin{equation}
Q_{i}=\frac{\hat{\partial}Q}{\hat{\partial}u^{i}},
\end{equation}
where $\hat{\partial}$ is taken on $\bP^{4}$, the $(3,0)$-form $\Omega$ can be written as
\begin{equation}
\Omega=Q_{\delta}^{-1}\prod_{i\neq\delta}\dd u^{i},
\end{equation}
where the coordinates on $X$ are taken as $u^i$ for $i\neq \delta$.\footnote{The $u^i$ give four coordinates on $\bP^4$. One of these, $i=\delta$, is singled out so that $u^\delta$ is defined implicitly by $Q=0$, with the remaining $u^i$, $i\neq\delta$, giving three coordinates on the hypersurface.} The volume form defined by $\Omega$ is then given by
\begin{equation}
\vol_{\Omega}=(-\ii)^{3}|Q_{\delta}|^{-2}\prod_{i\neq\delta}\text{d}u^{i}\wedge\prod_{j\neq\delta}\text{d}\bar{u}^{\bar{j}}.\label{eq: mu in X}
\end{equation}
The determinant of the metric $g_{a\bar{b}}$ on $X$, defined by $\KP[h]$, is then related to the metric $g_{i\bar{j}}$ on $\bP^{4}$ as
\begin{equation}
\det g_{a\bar{b}}=\frac{|Q|^{2}}{|Q_{\delta}|^{2}}\det\hat{g}_{i\bar{j}},
\end{equation}
where $|Q|^{2}\equiv\hat{g}^{i\bar{j}}Q_{i}\bar{Q}_{\bar{j}}$. Consequently, the ratio $v_{J}$ can be written as
\begin{equation}
v_{\omega}=|Q|^{2}\det\hat{g}^{i\bar{j}}.
\end{equation}
This is now expressed fully in terms of quantities computed on $\bP^{4}$. The determinant of the inverse metric on $\bP^{4}$, $\det\hat{g}^{i\bar{j}}$, can be written in terms of $\psi$ and consequently $h^{l}$. This yields
\begin{equation}
v_{\omega}=k^{-5}\psi^{-4}(\bar{Q}_{\bar{\beta}}\Psi^{\bar{\beta}\alpha}Q_{\alpha})\det\Psi_{\gamma\bar{\delta}},\label{eq:v_J definition}
\end{equation}
where the indices run over $\alpha=(i,c)$, and we have defined
\begin{equation}
\Psi_{\alpha\bar{\beta}}=h^{l}(\mathcal{Q}_{l})_{\alpha\bar{\beta}},\label{eq: vector notation Psi}
\end{equation}
with
\begin{equation}
q_{c}^{I}=\rho^{I},\quad q_{i}^{I}=\hat{\partial}_{i}\rho^{I},\quad(\mathcal{Q}_{l})_{\alpha\bar{\beta}}=c_{I\bar{J}}^{l}q_{\alpha}^{I}\bar{q}_{\bar{\beta}}^{\bar{J}},\label{eq: vector notation Q etc}
\end{equation}
and $Q_{\alpha}=(Q_{i},0)$. This form of $v_{\omega}$ can be calculated at randomly sampled points on $X$, inserted into (\ref{functional}), and then integrated.\footnote{Points on $X$ should be sampled according to the exact Calabi--Yau measure $\vol_{\Omega}$. In practice, this can be achieved either by rejection sampling~\cite{Headrick:2009jz} or by sampling according to a known, auxiliary distribution and then weighting the points appropriately to recover $\vol_{\Omega}$~\cite{Douglas:2006rr}} One then minimises $E[h]$ over the parameters $h^{l}$ in order to find the best approximation to the honest Ricci-flat metric. In summary, the algorithm consists of the following steps:
\begin{enumerate}
\item \textbf{Calculate the basis}. For a given $k$, find the basis of invariant polynomials $\mathcal{P}_{l}$.
\item \textbf{Generate points}. Generate random points on $\bP^{4}$ distributed according to $\vol_{\Omega}$. This yields $N_{p}$ tuples of four homogeneous coordinates $u^{i}$ that label the points.
\item \textbf{Calculate data}. Calculate $Q_{i}$, $\mathcal{Q}_{\alpha\bar{\beta}}^{l}$ and the matrix $\Psi_{\alpha\bar{\beta}}$ at each point. Using (\ref{eq:v_J definition}), compute the value of $v_{\omega}$ at each point.
\item \textbf{Integrate}. Compute $E[h]$ by summing the point-wise values of $(1-v_{\omega})^{2}$.
\item \textbf{Minimise}. Minimise $E[h]$ as a function of $h^{l}$. The values of $h^{l}$ obtained define the ``best'' approximate Ricci-flat metric for the chosen value of $k$.
\end{enumerate}
This algorithm was implemented in \texttt{Mathematica} and available in the \texttt{fermat.m} package at \cite{fermat.m}. The resulting metrics have come to be known as ``optimal metrics'', since they give the best possible approximation to the exact Ricci-flat metric within the family of algebraic metrics. As investigated in detail in \cite{Headrick:2009jz}, the accuracy of these metrics scales exponentially with $k$, rather than as a polynomial of $k$ as is the case for Donaldson's balanced metric approach.

In phenomenologically realistic models, the relevant Calabi--Yau spaces, such as the Schoen manifold used in \cite{Braun:2005ux}, generally admit much smaller discrete symmetry groups. The basis of invariant polynomials is then much larger, leading to a minimisation problem for a very large number of parameters.

\subsection{Data set}

\noindent Our data set $\mathcal{D}=\{u^{i},Q_{i},\mathcal{Q}_{\alpha\bar{\beta}}^{l}\to\psi\}$ consists of a set of 5000 inputs and outputs calculated using \cite{fermat.m}. The inputs are given by the point-wise quantities that enter the calculation of the functional $E[h]$. The outputs are given by the point-wise values of $\psi$ that correspond to an ``optimal'' approximate Ricci-flat metric. The set $\mathcal{D}$ is then split into a training set $\mathcal{T}$ and a validation set $\mathcal{V}$, with the fraction of $\mathcal{D}$ used for the training set denoted by $\gamma\in[0,1]$.

The outputs are calculated using the $k=10$ optimal metric. At $k=10$, there are 38 invariant polynomials $\mathcal{P}_{l}$ for the Fermat quintic. For each sample of the data set, $\mathcal{D}$ contains 958 complex input values (or 1916 real input values) and 1 real output. The neural network takes the input and produces a real number, which can then be compared with the target value of $\psi$. In what follows, the target values of $\psi$ corresponding to the optimal metric are denoted by $\psi$, while the values computed by the neural network are labelled $\hpsi$.

\subsection{Neural network\label{Neural networks and their parameters}}

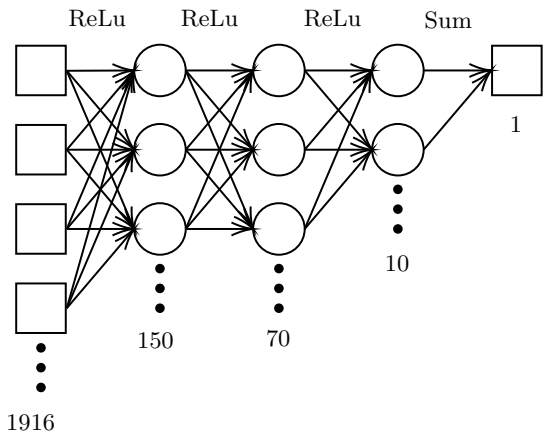
\begin{figure}[t]
\begin{tikzpicture}[x=0.75pt,y=0.75pt,yscale=-1,xscale=1]

\draw   (229,130) .. controls (229,122.82) and (234.82,117) .. (242,117) .. controls (249.18,117) and (255,122.82) .. (255,130) .. controls (255,137.18) and (249.18,143) .. (242,143) .. controls (234.82,143) and (229,137.18) .. (229,130) -- cycle ;
\draw   (229,170) .. controls (229,162.82) and (234.82,157) .. (242,157) .. controls (249.18,157) and (255,162.82) .. (255,170) .. controls (255,177.18) and (249.18,183) .. (242,183) .. controls (234.82,183) and (229,177.18) .. (229,170) -- cycle ;
\draw   (229,210) .. controls (229,202.82) and (234.82,197) .. (242,197) .. controls (249.18,197) and (255,202.82) .. (255,210) .. controls (255,217.18) and (249.18,223) .. (242,223) .. controls (234.82,223) and (229,217.18) .. (229,210) -- cycle ;
\draw   (289,130) .. controls (289,122.82) and (294.82,117) .. (302,117) .. controls (309.18,117) and (315,122.82) .. (315,130) .. controls (315,137.18) and (309.18,143) .. (302,143) .. controls (294.82,143) and (289,137.18) .. (289,130) -- cycle ;
\draw   (289,170) .. controls (289,162.82) and (294.82,157) .. (302,157) .. controls (309.18,157) and (315,162.82) .. (315,170) .. controls (315,177.18) and (309.18,183) .. (302,183) .. controls (294.82,183) and (289,177.18) .. (289,170) -- cycle ;
\draw   (289,210) .. controls (289,202.82) and (294.82,197) .. (302,197) .. controls (309.18,197) and (315,202.82) .. (315,210) .. controls (315,217.18) and (309.18,223) .. (302,223) .. controls (294.82,223) and (289,217.18) .. (289,210) -- cycle ;
\draw   (349,130) .. controls (349,122.82) and (354.82,117) .. (362,117) .. controls (369.18,117) and (375,122.82) .. (375,130) .. controls (375,137.18) and (369.18,143) .. (362,143) .. controls (354.82,143) and (349,137.18) .. (349,130) -- cycle ;
\draw   (349,170) .. controls (349,162.82) and (354.82,157) .. (362,157) .. controls (369.18,157) and (375,162.82) .. (375,170) .. controls (375,177.18) and (369.18,183) .. (362,183) .. controls (354.82,183) and (349,177.18) .. (349,170) -- cycle ;
\draw    (195,130.16) -- (227,130.01) ;
\draw [shift={(229,130)}, rotate = 539.72] [color={rgb, 255:red, 0; green, 0; blue, 0 }  ][line width=0.75]    (10.93,-3.29) .. controls (6.95,-1.4) and (3.31,-0.3) .. (0,0) .. controls (3.31,0.3) and (6.95,1.4) .. (10.93,3.29)   ;
\draw    (195,130.16) -- (227.7,168.48) ;
\draw [shift={(229,170)}, rotate = 229.52] [color={rgb, 255:red, 0; green, 0; blue, 0 }  ][line width=0.75]    (10.93,-3.29) .. controls (6.95,-1.4) and (3.31,-0.3) .. (0,0) .. controls (3.31,0.3) and (6.95,1.4) .. (10.93,3.29)   ;
\draw    (195,130.16) -- (228.22,208.16) ;
\draw [shift={(229,210)}, rotate = 246.93] [color={rgb, 255:red, 0; green, 0; blue, 0 }  ][line width=0.75]    (10.93,-3.29) .. controls (6.95,-1.4) and (3.31,-0.3) .. (0,0) .. controls (3.31,0.3) and (6.95,1.4) .. (10.93,3.29)   ;
\draw    (195,170) -- (227.7,131.52) ;
\draw [shift={(229,130)}, rotate = 490.36] [color={rgb, 255:red, 0; green, 0; blue, 0 }  ][line width=0.75]    (10.93,-3.29) .. controls (6.95,-1.4) and (3.31,-0.3) .. (0,0) .. controls (3.31,0.3) and (6.95,1.4) .. (10.93,3.29)   ;
\draw    (195,170) -- (227,170) ;
\draw [shift={(229,170)}, rotate = 180] [color={rgb, 255:red, 0; green, 0; blue, 0 }  ][line width=0.75]    (10.93,-3.29) .. controls (6.95,-1.4) and (3.31,-0.3) .. (0,0) .. controls (3.31,0.3) and (6.95,1.4) .. (10.93,3.29)   ;
\draw    (195,170) -- (227.7,208.48) ;
\draw [shift={(229,210)}, rotate = 229.64] [color={rgb, 255:red, 0; green, 0; blue, 0 }  ][line width=0.75]    (10.93,-3.29) .. controls (6.95,-1.4) and (3.31,-0.3) .. (0,0) .. controls (3.31,0.3) and (6.95,1.4) .. (10.93,3.29)   ;
\draw    (195,210) -- (227,210) ;
\draw [shift={(229,210)}, rotate = 180] [color={rgb, 255:red, 0; green, 0; blue, 0 }  ][line width=0.75]    (10.93,-3.29) .. controls (6.95,-1.4) and (3.31,-0.3) .. (0,0) .. controls (3.31,0.3) and (6.95,1.4) .. (10.93,3.29)   ;
\draw    (195,210) -- (228.22,131.84) ;
\draw [shift={(229,130)}, rotate = 473.03] [color={rgb, 255:red, 0; green, 0; blue, 0 }  ][line width=0.75]    (10.93,-3.29) .. controls (6.95,-1.4) and (3.31,-0.3) .. (0,0) .. controls (3.31,0.3) and (6.95,1.4) .. (10.93,3.29)   ;
\draw    (195,210) -- (227.7,171.52) ;
\draw [shift={(229,170)}, rotate = 490.36] [color={rgb, 255:red, 0; green, 0; blue, 0 }  ][line width=0.75]    (10.93,-3.29) .. controls (6.95,-1.4) and (3.31,-0.3) .. (0,0) .. controls (3.31,0.3) and (6.95,1.4) .. (10.93,3.29)   ;
\draw    (195,250) -- (227.7,211.52) ;
\draw [shift={(229,210)}, rotate = 490.36] [color={rgb, 255:red, 0; green, 0; blue, 0 }  ][line width=0.75]    (10.93,-3.29) .. controls (6.95,-1.4) and (3.31,-0.3) .. (0,0) .. controls (3.31,0.3) and (6.95,1.4) .. (10.93,3.29)   ;
\draw    (195,250) -- (228.22,171.84) ;
\draw [shift={(229,170)}, rotate = 473.03] [color={rgb, 255:red, 0; green, 0; blue, 0 }  ][line width=0.75]    (10.93,-3.29) .. controls (6.95,-1.4) and (3.31,-0.3) .. (0,0) .. controls (3.31,0.3) and (6.95,1.4) .. (10.93,3.29)   ;
\draw    (195,250) -- (228.45,131.92) ;
\draw [shift={(229,130)}, rotate = 465.82] [color={rgb, 255:red, 0; green, 0; blue, 0 }  ][line width=0.75]    (10.93,-3.29) .. controls (6.95,-1.4) and (3.31,-0.3) .. (0,0) .. controls (3.31,0.3) and (6.95,1.4) .. (10.93,3.29)   ;
\draw   (169.5,237.5) -- (194.5,237.5) -- (194.5,262.5) -- (169.5,262.5) -- cycle ;
\draw   (169.5,197.5) -- (194.5,197.5) -- (194.5,222.5) -- (169.5,222.5) -- cycle ;
\draw   (169.5,157.5) -- (194.5,157.5) -- (194.5,182.5) -- (169.5,182.5) -- cycle ;
\draw   (169.5,117.5) -- (194.5,117.5) -- (194.5,142.5) -- (169.5,142.5) -- cycle ;
\draw   (409.5,117.5) -- (434.5,117.5) -- (434.5,142.5) -- (409.5,142.5) -- cycle ;
\draw    (255,210) -- (287,210) ;
\draw [shift={(289,210)}, rotate = 180] [color={rgb, 255:red, 0; green, 0; blue, 0 }  ][line width=0.75]    (10.93,-3.29) .. controls (6.95,-1.4) and (3.31,-0.3) .. (0,0) .. controls (3.31,0.3) and (6.95,1.4) .. (10.93,3.29)   ;
\draw    (255,210) -- (287.7,171.52) ;
\draw [shift={(289,170)}, rotate = 490.36] [color={rgb, 255:red, 0; green, 0; blue, 0 }  ][line width=0.75]    (10.93,-3.29) .. controls (6.95,-1.4) and (3.31,-0.3) .. (0,0) .. controls (3.31,0.3) and (6.95,1.4) .. (10.93,3.29)   ;
\draw    (255,210) -- (288.22,131.84) ;
\draw [shift={(289,130)}, rotate = 473.03] [color={rgb, 255:red, 0; green, 0; blue, 0 }  ][line width=0.75]    (10.93,-3.29) .. controls (6.95,-1.4) and (3.31,-0.3) .. (0,0) .. controls (3.31,0.3) and (6.95,1.4) .. (10.93,3.29)   ;
\draw    (255,170) -- (287,170) ;
\draw [shift={(289,170)}, rotate = 180] [color={rgb, 255:red, 0; green, 0; blue, 0 }  ][line width=0.75]    (10.93,-3.29) .. controls (6.95,-1.4) and (3.31,-0.3) .. (0,0) .. controls (3.31,0.3) and (6.95,1.4) .. (10.93,3.29)   ;
\draw    (255,170) -- (287.7,131.52) ;
\draw [shift={(289,130)}, rotate = 490.36] [color={rgb, 255:red, 0; green, 0; blue, 0 }  ][line width=0.75]    (10.93,-3.29) .. controls (6.95,-1.4) and (3.31,-0.3) .. (0,0) .. controls (3.31,0.3) and (6.95,1.4) .. (10.93,3.29)   ;
\draw    (255,170) -- (287.7,208.48) ;
\draw [shift={(289,210)}, rotate = 229.64] [color={rgb, 255:red, 0; green, 0; blue, 0 }  ][line width=0.75]    (10.93,-3.29) .. controls (6.95,-1.4) and (3.31,-0.3) .. (0,0) .. controls (3.31,0.3) and (6.95,1.4) .. (10.93,3.29)   ;
\draw    (255,130) -- (287,130) ;
\draw [shift={(289,130)}, rotate = 180] [color={rgb, 255:red, 0; green, 0; blue, 0 }  ][line width=0.75]    (10.93,-3.29) .. controls (6.95,-1.4) and (3.31,-0.3) .. (0,0) .. controls (3.31,0.3) and (6.95,1.4) .. (10.93,3.29)   ;
\draw    (255,130) -- (287.7,168.48) ;
\draw [shift={(289,170)}, rotate = 229.64] [color={rgb, 255:red, 0; green, 0; blue, 0 }  ][line width=0.75]    (10.93,-3.29) .. controls (6.95,-1.4) and (3.31,-0.3) .. (0,0) .. controls (3.31,0.3) and (6.95,1.4) .. (10.93,3.29)   ;
\draw    (255,130) -- (288.22,208.16) ;
\draw [shift={(289,210)}, rotate = 246.97] [color={rgb, 255:red, 0; green, 0; blue, 0 }  ][line width=0.75]    (10.93,-3.29) .. controls (6.95,-1.4) and (3.31,-0.3) .. (0,0) .. controls (3.31,0.3) and (6.95,1.4) .. (10.93,3.29)   ;
\draw    (315,170) -- (347,170) ;
\draw [shift={(349,170)}, rotate = 180] [color={rgb, 255:red, 0; green, 0; blue, 0 }  ][line width=0.75]    (10.93,-3.29) .. controls (6.95,-1.4) and (3.31,-0.3) .. (0,0) .. controls (3.31,0.3) and (6.95,1.4) .. (10.93,3.29)   ;
\draw    (315,130) -- (347,130) ;
\draw [shift={(349,130)}, rotate = 180] [color={rgb, 255:red, 0; green, 0; blue, 0 }  ][line width=0.75]    (10.93,-3.29) .. controls (6.95,-1.4) and (3.31,-0.3) .. (0,0) .. controls (3.31,0.3) and (6.95,1.4) .. (10.93,3.29)   ;
\draw    (315,130) -- (347.7,168.48) ;
\draw [shift={(349,170)}, rotate = 229.64] [color={rgb, 255:red, 0; green, 0; blue, 0 }  ][line width=0.75]    (10.93,-3.29) .. controls (6.95,-1.4) and (3.31,-0.3) .. (0,0) .. controls (3.31,0.3) and (6.95,1.4) .. (10.93,3.29)   ;
\draw    (315,170) -- (347.7,131.52) ;
\draw [shift={(349,130)}, rotate = 490.36] [color={rgb, 255:red, 0; green, 0; blue, 0 }  ][line width=0.75]    (10.93,-3.29) .. controls (6.95,-1.4) and (3.31,-0.3) .. (0,0) .. controls (3.31,0.3) and (6.95,1.4) .. (10.93,3.29)   ;
\draw    (375,130) -- (407,130) ;
\draw [shift={(409,130)}, rotate = 180] [color={rgb, 255:red, 0; green, 0; blue, 0 }  ][line width=0.75]    (10.93,-3.29) .. controls (6.95,-1.4) and (3.31,-0.3) .. (0,0) .. controls (3.31,0.3) and (6.95,1.4) .. (10.93,3.29)   ;
\draw    (375,170) -- (407.7,131.52) ;
\draw [shift={(409,130)}, rotate = 490.36] [color={rgb, 255:red, 0; green, 0; blue, 0 }  ][line width=0.75]    (10.93,-3.29) .. controls (6.95,-1.4) and (3.31,-0.3) .. (0,0) .. controls (3.31,0.3) and (6.95,1.4) .. (10.93,3.29)   ;
\draw    (315,210) -- (347.7,171.52) ;
\draw [shift={(349,170)}, rotate = 490.36] [color={rgb, 255:red, 0; green, 0; blue, 0 }  ][line width=0.75]    (10.93,-3.29) .. controls (6.95,-1.4) and (3.31,-0.3) .. (0,0) .. controls (3.31,0.3) and (6.95,1.4) .. (10.93,3.29)   ;
\draw    (315,210) -- (348.22,131.84) ;
\draw [shift={(349,130)}, rotate = 473.03] [color={rgb, 255:red, 0; green, 0; blue, 0 }  ][line width=0.75]    (10.93,-3.29) .. controls (6.95,-1.4) and (3.31,-0.3) .. (0,0) .. controls (3.31,0.3) and (6.95,1.4) .. (10.93,3.29)   ;
\draw  [color={rgb, 255:red, 0; green, 0; blue, 0 }  ,draw opacity=1 ][fill={rgb, 255:red, 0; green, 0; blue, 0 }  ,fill opacity=1 ] (180.09,269.97) .. controls (180.11,268.92) and (180.97,268.08) .. (182.03,268.09) .. controls (183.08,268.11) and (183.92,268.97) .. (183.91,270.03) .. controls (183.89,271.08) and (183.03,271.92) .. (181.97,271.91) .. controls (180.92,271.89) and (180.08,271.03) .. (180.09,269.97) -- cycle ;
\draw  [color={rgb, 255:red, 0; green, 0; blue, 0 }  ,draw opacity=1 ][fill={rgb, 255:red, 0; green, 0; blue, 0 }  ,fill opacity=1 ] (180.09,279.97) .. controls (180.11,278.92) and (180.97,278.08) .. (182.03,278.09) .. controls (183.08,278.11) and (183.92,278.97) .. (183.91,280.03) .. controls (183.89,281.08) and (183.03,281.92) .. (181.97,281.91) .. controls (180.92,281.89) and (180.08,281.03) .. (180.09,279.97) -- cycle ;
\draw  [color={rgb, 255:red, 0; green, 0; blue, 0 }  ,draw opacity=1 ][fill={rgb, 255:red, 0; green, 0; blue, 0 }  ,fill opacity=1 ] (180.09,289.97) .. controls (180.11,288.92) and (180.97,288.08) .. (182.03,288.09) .. controls (183.08,288.11) and (183.92,288.97) .. (183.91,290.03) .. controls (183.89,291.08) and (183.03,291.92) .. (181.97,291.91) .. controls (180.92,291.89) and (180.08,291.03) .. (180.09,289.97) -- cycle ;
\draw  [color={rgb, 255:red, 0; green, 0; blue, 0 }  ,draw opacity=1 ][fill={rgb, 255:red, 0; green, 0; blue, 0 }  ,fill opacity=1 ] (240.09,229.97) .. controls (240.11,228.92) and (240.97,228.08) .. (242.03,228.09) .. controls (243.08,228.11) and (243.92,228.97) .. (243.91,230.03) .. controls (243.89,231.08) and (243.03,231.92) .. (241.97,231.91) .. controls (240.92,231.89) and (240.08,231.03) .. (240.09,229.97) -- cycle ;
\draw  [color={rgb, 255:red, 0; green, 0; blue, 0 }  ,draw opacity=1 ][fill={rgb, 255:red, 0; green, 0; blue, 0 }  ,fill opacity=1 ] (240.09,239.97) .. controls (240.11,238.92) and (240.97,238.08) .. (242.03,238.09) .. controls (243.08,238.11) and (243.92,238.97) .. (243.91,240.03) .. controls (243.89,241.08) and (243.03,241.92) .. (241.97,241.91) .. controls (240.92,241.89) and (240.08,241.03) .. (240.09,239.97) -- cycle ;
\draw  [color={rgb, 255:red, 0; green, 0; blue, 0 }  ,draw opacity=1 ][fill={rgb, 255:red, 0; green, 0; blue, 0 }  ,fill opacity=1 ] (240.09,249.97) .. controls (240.11,248.92) and (240.97,248.08) .. (242.03,248.09) .. controls (243.08,248.11) and (243.92,248.97) .. (243.91,250.03) .. controls (243.89,251.08) and (243.03,251.92) .. (241.97,251.91) .. controls (240.92,251.89) and (240.08,251.03) .. (240.09,249.97) -- cycle ;
\draw  [color={rgb, 255:red, 0; green, 0; blue, 0 }  ,draw opacity=1 ][fill={rgb, 255:red, 0; green, 0; blue, 0 }  ,fill opacity=1 ] (360.09,189.97) .. controls (360.11,188.92) and (360.97,188.08) .. (362.03,188.09) .. controls (363.08,188.11) and (363.92,188.97) .. (363.91,190.03) .. controls (363.89,191.08) and (363.03,191.92) .. (361.97,191.91) .. controls (360.92,191.89) and (360.08,191.03) .. (360.09,189.97) -- cycle ;
\draw  [color={rgb, 255:red, 0; green, 0; blue, 0 }  ,draw opacity=1 ][fill={rgb, 255:red, 0; green, 0; blue, 0 }  ,fill opacity=1 ] (360.09,209.97) .. controls (360.11,208.92) and (360.97,208.08) .. (362.03,208.09) .. controls (363.08,208.11) and (363.92,208.97) .. (363.91,210.03) .. controls (363.89,211.08) and (363.03,211.92) .. (361.97,211.91) .. controls (360.92,211.89) and (360.08,211.03) .. (360.09,209.97) -- cycle ;
\draw  [color={rgb, 255:red, 0; green, 0; blue, 0 }  ,draw opacity=1 ][fill={rgb, 255:red, 0; green, 0; blue, 0 }  ,fill opacity=1 ] (360.09,199.97) .. controls (360.11,198.92) and (360.97,198.08) .. (362.03,198.09) .. controls (363.08,198.11) and (363.92,198.97) .. (363.91,200.03) .. controls (363.89,201.08) and (363.03,201.92) .. (361.97,201.91) .. controls (360.92,201.89) and (360.08,201.03) .. (360.09,199.97) -- cycle ;
\draw  [color={rgb, 255:red, 0; green, 0; blue, 0 }  ,draw opacity=1 ][fill={rgb, 255:red, 0; green, 0; blue, 0 }  ,fill opacity=1 ] (300.09,229.97) .. controls (300.11,228.92) and (300.97,228.08) .. (302.03,228.09) .. controls (303.08,228.11) and (303.92,228.97) .. (303.91,230.03) .. controls (303.89,231.08) and (303.03,231.92) .. (301.97,231.91) .. controls (300.92,231.89) and (300.08,231.03) .. (300.09,229.97) -- cycle ;
\draw  [color={rgb, 255:red, 0; green, 0; blue, 0 }  ,draw opacity=1 ][fill={rgb, 255:red, 0; green, 0; blue, 0 }  ,fill opacity=1 ] (300.09,239.97) .. controls (300.11,238.92) and (300.97,238.08) .. (302.03,238.09) .. controls (303.08,238.11) and (303.92,238.97) .. (303.91,240.03) .. controls (303.89,241.08) and (303.03,241.92) .. (301.97,241.91) .. controls (300.92,241.89) and (300.08,241.03) .. (300.09,239.97) -- cycle ;
\draw  [color={rgb, 255:red, 0; green, 0; blue, 0 }  ,draw opacity=1 ][fill={rgb, 255:red, 0; green, 0; blue, 0 }  ,fill opacity=1 ] (300.09,249.97) .. controls (300.11,248.92) and (300.97,248.08) .. (302.03,248.09) .. controls (303.08,248.11) and (303.92,248.97) .. (303.91,250.03) .. controls (303.89,251.08) and (303.03,251.92) .. (301.97,251.91) .. controls (300.92,251.89) and (300.08,251.03) .. (300.09,249.97) -- cycle ;

\draw (163,302) node [anchor=north west][inner sep=0.75pt]  [font=\small] [align=left] {1916};
\draw (229,261) node [anchor=north west][inner sep=0.75pt]  [font=\small] [align=left] {150};
\draw (294,260) node [anchor=north west][inner sep=0.75pt]  [font=\small] [align=left] {70};
\draw (354,222) node [anchor=north west][inner sep=0.75pt]  [font=\small] [align=left] {10};
\draw (417,152) node [anchor=north west][inner sep=0.75pt]  [font=\small] [align=left] {1};
\draw (194.37,98) node [anchor=north west][inner sep=0.75pt]  [font=\small] [align=left] {ReLu};
\draw (251,98) node [anchor=north west][inner sep=0.75pt]  [font=\small] [align=left] {ReLu};
\draw (313,98) node [anchor=north west][inner sep=0.75pt]  [font=\small] [align=left] {ReLu};
\draw (374,99) node [anchor=north west][inner sep=0.75pt]  [font=\small] [align=left] {Sum};
\end{tikzpicture}\caption{The neural network contains five layers of widths $(1916,150,70,10,1)$. The input layer takes values from $\{u^{i},Q_{i},\mathcal{Q}_{\alpha\bar{\beta}}^{l}\}$ and the output node encodes $\protect\hpsi$. A drop-out rate of $0.01$ in the first hidden layer is not shown.}
\end{figure}

\noindent In order to predict point-wise values of $\hpsi$ from $\{u^{i},Q_{i},\mathcal{Q}_{\alpha\bar{\beta}}^{l}\}$, we use a five-layer neural network with widths $(1916,150,70,10,1)$ implemented in \texttt{Python} using the \texttt{TensorFlow} library~\cite{tensorflow2015-whitepaper}. At each node of the network, the weights and biases together with the activation functions of each layer determine the flow of information through the network. We use the Rectified Linear Unit (ReLU) activation function for all layers but the last one. The network then essentially acts as a function, mapping 1916 arguments onto the reals. The power and flexibility of the network comes from the non-linearity of the ReLu activation function, which allows it to capture the complex relation between the input and output data; in our case an integration over many points and the minimisation of a functional. In addition, we found a $0.01$ drop-out rate in the first hidden layer was useful to ensure the network learned global structures instead of the specific features of the training data (which might not generalise to unseen inputs). This randomly deactivates nodes during each pass over the data and helps to avoid overfitting, a phenomenon where the network's performance improves with more training data but underperforms when tested on validation data.

During the training phase, the internal parameters of the network (the weights and biases of each node) are adjusted to minimise a loss function, which we chose as the mean absolute percentage error (MAPE), given by
\begin{equation}
\text{MAPE}=100\cdot\frac{1}{n}\sum_{p=1}^{n}\frac{|\psi-\hpsi|_{p}}{\psi|_{p}},\label{eq: mape def}
\end{equation}
where $n$ is the number of points considered in each training round, and the index $p$ on $\psi|_{p}$ indicates the value of $\psi$ at the point $p$.

The training proceeds as follows: given a batch of samples drawn randomly from the training set $\mathcal{T}$, the predicted outputs $\hpsi$ are computed by the network and compared to target values of $\psi$ using the MAPE. The parameters of the network are then adjusted via gradient descent to reduce the MAPE for all points in the batch. This is repeated with new batches until all data in $\mathcal{T}$ has been presented to the network. This forms one ``epoch''. The network is then trained over multiple epochs, adjusting its parameters each time until the MAPE is reduced to an acceptable value. We trained our network for $100$ epochs, with batches of $70$ points.

\section{Machine learning the Kähler potential\label{sec:h11}}

\subsection{A single training}

\begin{figure}
\includegraphics[trim = 1cm 0.6cm 1cm 0.3cm]{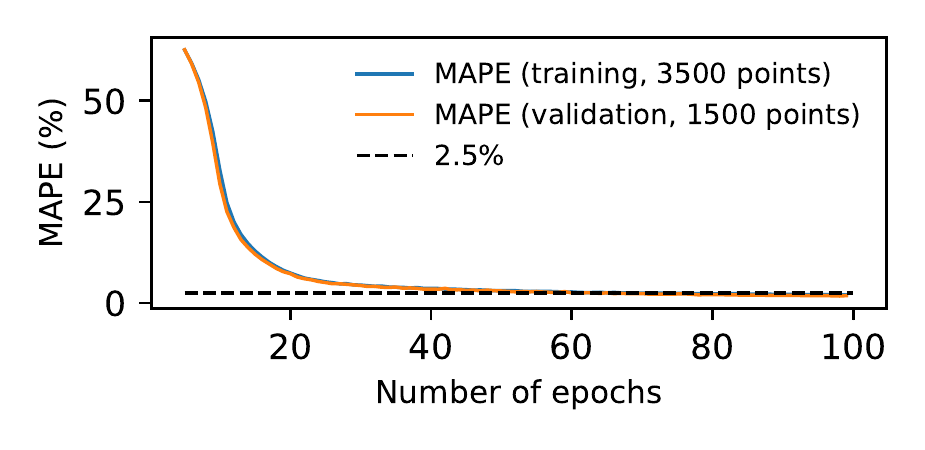} \caption{The loss for both training and validation sets after each epoch. Training lasted for 100 epochs in total. As a guide, the $2.5\%$ line is also shown.}
\label{fig: training mape}
\end{figure}

\noindent Training the network with $\gamma=0.7$ (i.e.~3500 points for training and 2500 for validation), we calculate the value of the loss function (MAPE) after each epoch and plot this in Fig.~\ref{fig: training mape} for both the training and validation sets. The loss clearly decreases as training proceeds. After 100 epochs, the discrepancy between the predictions of the network, $\hpsi$, and the values corresponding to the optimal $k=10$ metric, $\psi$, drops to $1.6\%$, showing that the network can indeed learn an accurate representation of the approximately Ricci-flat Kähler potential.

Curiously, the loss on the training set remains $1\%$ above the loss on the validation set. This behaviour can be traced to the 0.01 drop-out rate, where, on average, about 1.5 nodes are removed during each pass. This highlights the subtlety of neural networks, where the removal of a single node can significantly alter or improve performance. This result confirms that our neural network is not overfitting to features in the training set and can extrapolate well to unseen data.

\begin{figure}
\includegraphics[trim = 1cm 0.6cm 1cm 0.3cm]{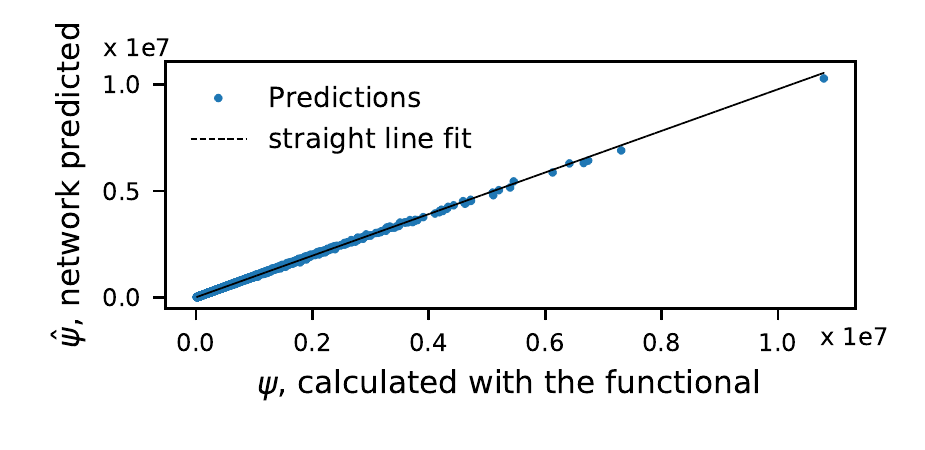} \caption{The point-wise predictions $\protect\hpsi|_{p}$ of the trained network plotted against the target values $\psi|_{p}$ calculated using the optimal $k=10$ metric for 10000 previously unseen points. We also show a linear fit with slope $1.012\pm0.002$, where the uncertainties represent standard deviation errors on the parameters.}
\label{fig: fitted data} 
\end{figure}

Using this trained neural network on a new set of 10000 input and output samples (akin to an unseen validation set), one can obtain $\hpsi$ from the trained neural network for these 10000 points in roughly one second. In Fig.~\ref{fig: fitted data}, we present the predicted outputs $\hpsi$ for these 10000 points against the corresponding values of $\psi$ calculated using the optimal $k=10$ metric. Using a linear fit, we find a slope of $1.012\pm0.002$, confirming the high accuracy of the predicted values of $\hat{\psi}$. This single training helps to verify that our architecture does not overfit local features of the training set.

\subsection{Training over multiple \texorpdfstring{$\gamma$}{gamma}}

\begin{figure}
\includegraphics[trim = 1cm 0.6cm 1cm 0.3cm]{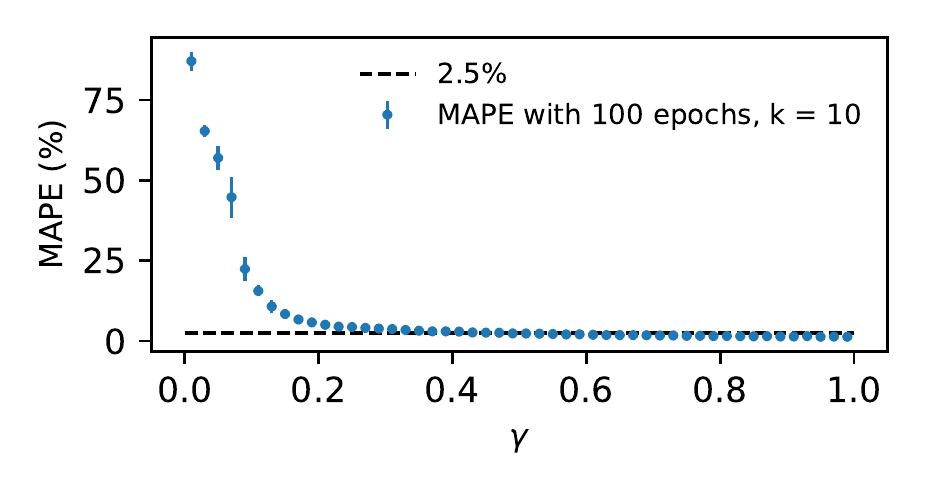} \caption{The loss for validation sets averaged over ten runs for each training fraction $\gamma$. The error bars represent the standard error on the average. As a guide, the $2.5\%$ line is also shown.}
\label{fig: learning mape} 
\end{figure}

\noindent In order to assess how the performance of the network is affected by the size of the training set, we compute the learning curve (the value of the MAPE) for varying training set fractions, $\gamma$. For each $\gamma$, the network is trained from scratch using the dataset $\mathcal{D}$ described above. The learning curve, averaged over ten independent realisations of the network, is shown in Fig.~\ref{fig: learning mape} together with the corresponding standard error on the mean.

\begin{figure}
\includegraphics[trim = 1cm 0.6cm 1cm 0.3cm]{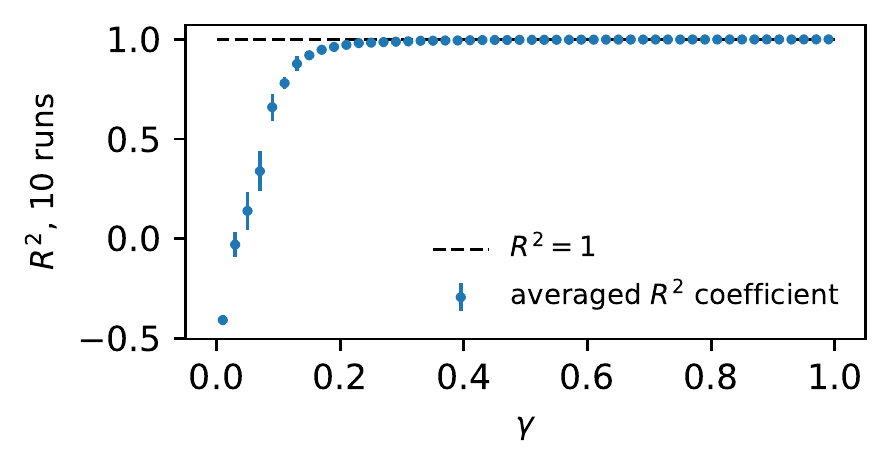} \caption{The value of the R-squared coefficient, $R^{2}$, for validation sets averaged over ten runs computed for varying training fraction, $\gamma$. The error bars represent the standard error on the mean. We overlay the $R^{2}=1$ line, which represents a perfect fit.}
\label{fig: learning r_2} 
\end{figure}

The mean value of the loss (itself a mean percentage error) clearly decreases as the amount of data seen during training increases, i.e.~the network is learning from the data given. The error on the averaged loss decreases as $\gamma$ increases, showing that the network is learning reliably from one training to the next. Specifically, when learning from $\gamma=0.4$ of the data (i.e.~2000 points), the MAPE is already below $3\%$. It reduces to $2.5\%\pm0.1\%$ at $\gamma=0.46$ (i.e.~having seen $2300$ points). For $\gamma=0.78$ ($3900$ points), the error drops to $1.5\%\pm0.07\%$. This demonstrates that our relatively simple network architecture is able to learn significant features of our data set, encoding the optimal $k=10$ metric to an excellent accuracy that varies negligibly between different trainings. To verify this, we compute the ten-run average of the R-squared coefficient,
\begin{equation}
R^{2}(\psi,\hat{\psi})\equiv1-\frac{\sum_{p}(\psi-\hpsi)_{p}^{2}}{\sum_{p}(\psi|_{p}-\bar{\psi})^{2}},\label{eq: r2}
\end{equation}
where $\bar{\psi}$ is the statistical mean of all $\psi|_{p}$~\cite{He:2018jtw}. This is plotted in Fig.~\ref{fig: learning r_2} for the validation set for each choice of training fraction, $\gamma$. The $R^{2}$ coefficient reaches $1$ around $\gamma=0.4$, further confirming that the network is able to learn significant underlying features of the data with only $2000$ points. The exact values are $R^{2}=0.996\pm0.001$ for $\gamma=0.46$ and $R^{2}=0.998\pm0.0003$ for $\gamma=0.78$, again confirming the high accuracy of the results produced with the neural network. 

In summary, the error on the ten-run averaged MAPE and $R^{2}$ coefficients shown in Figs.~\ref{fig: learning mape} and \ref{fig: learning r_2} decreases quickly with increasing $\gamma$: predictions are poor for $\gamma\apprle0.2$, where the standard error for ten-run averages are largest; after stabilising at $\gamma\apprge0.2$, the values of $\hpsi$ produced are of approximately constant accuracy.

On a personal laptop, generating the data set $\mathcal{D}$ using the \texttt{fermat.m} package took 16 minutes, of which 10 minutes were needed to calculate the input data and 6 minutes to perform the minimisation and obtain the values of $\psi$. In comparison, our neural network architecture completes training in under 2 minutes and predicts values of $\hpsi$ with approximately $1.5\%$ error in under a second. The network is capable of learning the features underlying the relationship between $\psi$ and the input data with great accuracy. In particular, it predicts $\psi$ with less than $2.5\%$ error after having seen only 2300 data points. Training the network on 3500 points (with error around $1.5\%$) takes 2 minutes and obtaining the predictions for 10000 additional points takes under 2 seconds. This is a considerable gain from the $6$ minutes required to compute the $5000$ values of $\psi$ with the minimisation. The simple network structure detailed in Section \ref{Neural networks and their parameters} is thus capable of reproducing the relationship between $\{u^{i},Q_{i},\mathcal{Q}_{\alpha\bar{\beta}}^{l}\}$, i.e.~the coordinates and invariant sections, and $\psi$, i.e.~the exponential of the Kähler potential.

\begin{acknowledgments}
\noindent AA is supported by the European Union's Horizon 2020 research and innovation program under the Marie Skłodowska-Curie grant agreement No.~838776. LC thanks Adam Chalabi for useful discussions. YHH would like to thank STFC for grant ST/J00037X/1. BAO is supported in part by both the research grant DOE No.~DESC0007901 and SAS Account 020-0188-2-010202-6603-0338.
\end{acknowledgments}

\bibliographystyle{utphys}
\bibliography{main}

\end{document}